\documentstyle[11pt,aas2pp4]{article}

\slugcomment{Submitted to ApJ Letters}

\lefthead{Gieren et al.}
\righthead{Distance to the LMC}

\begin{document}

\title{A Direct Distance to the LMC Cepheid HV\,12198 from the
Infrared Surface Brightness Technique}

\author{W. P. Gieren\altaffilmark{1,2}}
\affil{Universidad de Concepci\'on, Departamento de F\'{\i}sica, 
Casilla 160-C, Concepci\'on, Chile}

\author{J. Storm\altaffilmark{1}}
\affil{Astrophysikalisches Institut Potsdam, An der Sternwarte 16, D-14482 
Potsdam, 
Germany}

\author{P. Fouqu\'e\altaffilmark{1}}
\affil{Observatoire de Paris, Section de Meudon, DESPA F-92195 Meudon Cedex, 
France\\
European Southern Observatory, Casilla 19001, Santiago 19, Chile}

\author{R.E. Mennickent}
\affil{Universidad de Concepci\'on, Departamento de F\'{\i}sica, 
Casilla 160-C, Concepci\'on, Chile}

\and

\author{M. G\'omez\altaffilmark{2}}

\affil{P. Universidad Cat\'olica de Chile, Departamento de Astronom\'{\i}a y 
Astrof\'{\i}sica, 
Casilla 104, Santiago 22, Chile}


\altaffiltext{1}{Visiting astronomer, European Southern Observatory, La Silla, 
Chile } 
\altaffiltext{2}{Visiting astronomer, Las Campanas Observatory, Carnegie  
Institution of Washington, Chile}


\begin{abstract}
We report on a first application of the infrared surface brightness  
technique on a Cepheid in the Large Magellanic Cloud, the variable 
HV 12198 in the young globular 
cluster NGC\,1866. From this one star, we determine a distance modulus of 18.42 
$\pm$ 0.10 (random and systematic uncertainty) to the cluster.  
When the results on further member Cepheids in NGC\,1866 become 
available, we expect to derive the distance to the LMC with a $\pm$ 3-4 percent 
accuracy, including systematic errors, from this technique.
\end{abstract}




%

\section{Introduction}
 
 The determination of the distance to our neighboring galaxy, the Large Magellanic 
Cloud, is a problem of fundamental astrophysical importance. Objects of any kind can be 
studied in the LMC in greater detail than in any other galaxy, but in order to know 
luminosities and true linear dimensions, we have to know an accurate distance to these 
objects. For the extragalactic distance scale, the LMC plays a fundamental role because 
most of the extragalactic distance calibrators, like Cepheids, are tied to the distance of the 
LMC, either directly or indirectly. 
     Due to its prime astrophysical importance there has been a large effort to determine the 
distance to the LMC from many different objects, and with a variety of techniques. The 
current state of this subject has recently been reviewed comprehensively by Walker (1999). 
Currently, distance determinations for the LMC scatter between 18.2 and 18.7 mag, a very 
unsatisfactory situation which clearly indicates the presence of significant systematic errors 
in most of the methods. Therefore, any new technique which holds the promise of yielding 
a truly accurate distance to the LMC is of great astrophysical relevance.

\section{The infrared surface brightness technique}

Following an original idea of Welch (1994), Fouqu\'e~' \& Gieren (1997) have calibrated 
two versions of a technique which employs the V, V-K and K, J-K magnitude/color 
combinations, respectively, to derive the angular diameter  of a Cepheid variable, at any 
given phase of its pulsation cycle. This angular diameter can be combined with the 
instantaneous linear diameter of the pulsating star, at the same phase, which is derived from 
an integration of its observed radial velocity curve. Observing many such pairs of angular 
and linear diameters over the pulsation cycle, one can derive both, the mean radius, and the 
distance of the Cepheid variable from a simple regression analysis. The distance so derived 
is independent of the pulsation mode of the variable (although in practice application will 
be mostly restricted to fundamental mode pulsators with their larger light- and color 
amplitudes). An application of this infrared technique to a large number of Galactic 
Cepheids (\cite{gie97,gie98}) has shown that the combined random and 
systematic uncertainty of a Cepheid distance can be as low as $\sim$  $\pm$ 3 percent, if the data used 
in the analysis are of excellent quality. In these papers it was also shown that the distances 
from both versions of the method agree to better than 2 percent. A very important feature of 
the technique is its very low sensitivity to the assumed reddening, and to the Cepheid's 
metallicity. These features make the infrared surface brightness technique an excellent tool 
to derive direct, one-step Cepheid distances of very low systematic uncertainty to nearby 
galaxies, circumventing most of the problems due to uncertain absorption and metallicity 
corrections, and could provide the means to finally beat down the uncertainty of the LMC 
distance below the 3 percent level, which is scientifically so desirable.

\section{The LMC cluster NGC\,1866 and its cepheid population}

The ideal targets in the LMC to derive a very accurate Cepheid distance from the 
infrared surface brightness technique are the young globular clusters. Several of these 
objects contain considerable numbers of Cepheid variables whose individual distances can 
be averaged to determine accurate mean cluster distances. Also, an analysis of the 
dispersion among the distances of a number of member Cepheids of a given cluster will 
yield the ultimate test of the true capabilities of the technique.

Among the Cepheid-rich LMC clusters, the outstanding object is NGC\,1866. It was the 
first LMC cluster in which Cepheid variables were detected (\cite{arp67}), and 
later surveys for more Cepheid variables in regions closer to the cluster center (\cite{sto98,wel91}) turned up a total number of Cepheids in excess of 20 (which is 
about the same as all Cepheids in Galactic open clusters taken together!). Welch et al.\ 
(1991), in a pioneering study, were the first to collect radial velocity data for a significant 
fraction of the NGC\, 1866 Cepheids which indicated that all these objects (with the possible 
exception of 1 star) are indeed cluster members. A first Baade-Wesselink-type analysis to 
determine mean radii and distances for some of the outer cluster Cepheids was performed 
by Cot\'e' et al.\ (1991), using photometric data in the B and V bands. The distances derived in 
this work showed a large dispersion. Similar work using the visual surface brightness 
technique (\cite{gie93}) was carried out by Gieren, Richtler \& Hilker 
(1994), and led to an improved distance of the cluster. However, the accuracy of their result 
($\pm$ 0.2 mag) was still not good enough for a breakthrough, which had to await the advent of 
improved techniques calibrated in the infrared spectral range.

\section{New observations and data reduction}

Given the potential usefulness of NGC\,1866 to derive a very accurate Cepheid distance 
to the LMC with the infrared surface brightness technique, we started to collect the 
necessary data in 1995. Three types of data were obtained: CCD images through BVRI 
(Cousins) filters; near-infrared images in JK filters; and high resolution, low S/N Echelle 
spectra. All CCD images were obtained on the 1m Swope reflector of Las Campanas 
Observatory, during four different runs in 1994-1996. The near-infrared imaging was done 
between 1996 and 1998 on the Las Campanas 2.5 and 1m-telescopes using the respective 
infrared cameras, and on the ESO 2.2m telescope with the IRAC2 instrument. Spectra of 
six Cepheid variables in NGC 1866 were obtained with the photon-counting Echelle 
spectrograph on the 2.5m Dupont reflector of Las Campanas Observatory, during several 
runs between 1995 and 1996; the same instrument had been used before by Welch et 
al.\ (1991) to obtain their first-epoch radial velocity observations of these stars.
    
Photometry was done on the CCD data using the DoPHOT (\cite{sch93}) package. It was done differentially with respect to a large number of local 
comparison stars which were tied to Landolt standards observed on photometric nights. The 
infrared data were reduced using standard IRAF reduction packages and the photometry 
was performed using DoPHOT. 
     Radial velocities were extracted from the Echelle spectra using IRAF routines. For 
typical Cepheid V magnitudes close to 16 and a limit integration time per spectrum of 1800 
sec (set by the short periods of the variables), signal-to-noise ratios under typical seeing 
conditions were low, but high enough to extract radial velocities with standard errors of 1-2 
km/s. The star AT67A was observed as a local reference star following Welch et al.\ (1991). 
The final radial velocities were determined from the 3950-6050~\AA~ spectral range. A full 
description of the photometric and radial velocity reduction procedures will be given in 
follow-up papers (\cite{gie00,sto00}). In these papers, we will also 
present the new data.

\section{Distance and radius of the cepheid variable HV\,12198}

 At the present time, dataset and reductions are complete for the Cepheid HV\,12198 (star 
"g" on the finder chart given by Arp \& Thackeray 1967, located about 1.4 arcmin from the 
center of NGC\,1866). Judging from its asymmetric light curve and the amplitudes of its 
light and radial velocity curves, this variable is most likely a fundamental mode pulsator. A 
Fourier decomposition analysis of the light curves of all the NGC\,1866 Cepheids to 
determine their pulsation modes will be presented in a forthcoming paper. For the purpose 
of this Letter, it suffices to recall that our method of distance determination does not depend 
on the pulsation mode. The relatively red (B-V) color of HV\,12198 seems to be normal for 
the Cepheids in NGC\,1866 (\cite{gie00}) and not hinting at the possibility of 
blending with a luminous red star, which could affect our analysis in a systematic way. The 
period of HV\,12198 determined from our new and literature V data is 3.52279 days. 
We do not find evidence for a variable period for this star, as suggested in Cot\'e et al.\ (1991); 
rather, a constant period fits all the existing data sets very well (\cite{gie00}). In Figs.\ 
1-3, we show the radial velocity curve, and the V and K band light curves of HV\,12198 
obtained from our new data. The mean radial velocity of 299.8 $\pm$ 0.1 km/s (intrinsic error of 
the mean; including systematics, the uncertainty is 1-2 km/s) derived from our new data 
strongly supports cluster membership (see \cite{wel91}). It is identical to the value 
derived earlier by Welch et al., indicating that the systemic radial velocity has not changed, 
i.e.\ that HV\,12198 is not a spectroscopic binary.
     In order to apply the V, V-K version of the infrared surface brightness technique we had 
to determine the V-K color curve of HV\,12198. Since our V and K observations were not 
obtained simultaneously, we adopted the approach to fit a Fourier series to the K data to 
obtain the K values at the phases of the actual V observations. Given the constancy and 
accuracy of the period of the Cepheid, this procedure should not introduce a significant 
systematic uncertainty into the distance solution. The V and V-K observations were 
dereddened adopting $A_{V}=3.2 E_{B-V}$, $E_{V-K}=0.88 A_{V}$ (see 
\cite{fou97}), and $E_{B-V}$=0.07 
which is appropriate for the NGC\,1866 field (\cite{fea87}). We recall that 
the distance derived from the infrared surface brightness technique is very insensitive to 
uncertainties in these values (effect on the distance is less than 1 percent for an uncertainty 
of $\pm$ 0.04 in  $E_{B-V}$; see \cite{gie97}). Angular diameters were then 
calculated from relation (35) of Fouqu\'e' \& Gieren (1997) at the phases of the photometric 
observations, yielding the angular diameter curve of the Cepheid.
     A fourth order Fourier series fit gave an excellent representation of the observed radial 
velocities (including the Welch et al.\ 1991 data; see Fig. 1). From the fitted curve, the linear 
displacements at the phases of the V observations were determined adopting a p factor of 
1.374 appropriate to the period of HV\,12198 (\cite{gie93}). The 
displacements were combined with the angular diameters at the same phases to solve for the 
distance and mean radius of the variable from an inverse linear regression, as described and 
discussed in \cite{gie97}. This yields a distance of (48.34 $\pm$ 1.80) kpc and a mean 
radius of (32.9 $\pm$ 1.2) R$_{\odot}$. The corresponding diagram from which these solutions were 
derived is shown in Fig.\ 4. We tested the stability of this solution in different ways: by 
using the actual K data and smoothing the V data to construct the V-K color curve; and by 
using the actual V data and interpolating linearly between adjacent (in phase) K 
observations (no smoothing) to determine V-K at the phases of the V observations. The 
distance and radius values changed by less than 1 $\sigma$ when these different procedures to 
obtain the V-K data were adopted. Also, we verified that a different way to obtain the radial 
velocity curve (interpolating linearly between adjacent velocity observations) did not 
change the results by more than a few tenth of 1 $\sigma$. As a conclusion, our distance and radius 
result for HV\,12198 seems to be very robust with respect to the way the data are fitted for 
the analysis, which just reflects the fact that the photometric variations and the radial 
velocity curve are very well defined with the present data.
     The radius value of 32.9 R$_{\odot}$ is in very good agreement with the radius of 30.6 
R$_{\odot}$ predicted for a 3.5 day Cepheid by the  period-radius relation of Gieren, 
Fouqu\'e \& G\'omez (1998) calibrated on Galactic stars. This strengthens the recent finding of Gieren, 
Moffett \& Barnes (1999) that Cepheids of the Galaxy and both Magellanic Clouds appear 
to obey the same PR relation. It also implies that any dependence of the Cepheid period-
radius relation on metallicity must be small enough to escape detection, within the accuracy 
of the radius determinations.

\section{The distance to the LMC}

The distance of HV\,12198 found in this paper corresponds to a true distance modulus of 
(18.42 $\pm$ 0.08) mag for its host cluster NGC 1866. Taking into account a possible 
systematic error on the distance of $\pm$ 3 percent (\cite{gie97}), the total 
uncertainty on the distance modulus is $\pm$ 0.10 mag. This result is almost identical to the 
LMC distance modulus of 18.46 obtained by Gieren, Fouqu\'e \& G\'omez (1998)  from a 
comparison of LMC Cepheid PL relations in VIJHK passbands to the Galactic relation 
obtained from infrared surface brightness distances to some 30 Galactic Cepheids. Since 
this latter result has, in principle, a dependence on the poorly known metallicity effect on 
Cepheid luminosities, our current direct and metallicity-independent distance result for 
HV\,12198 suggests that the effect of metallicity on Cepheid absolute magnitudes in optical and 
near-infrared photometric bands is small and might even be zero. However, this is a 
preliminary conclusion, which can be checked and quantified once the distance results for  
the other NGC\,1866 Cepheids, and hence a more accurate LMC distance, become available. 
For the time being, let us also note that a slightly shorter distance modulus to the LMC than 
that of 18.5  favored by the HST Key Project team tends to bring the discrepant results on 
the distance of NGC\,4258 (\cite{her99,mao99}) into better agreement.
      The low $\sim$ $\pm$ 5 percent uncertainty of the current distance determination for one Cepheid 
in NGC\,1866 makes us optimistic that our study will achieve its principal goal to determine 
a Cepheid-based distance to the LMC which will have lower random and systematic 
uncertainty than any previous determination with Cepheids, and will mean true progress 
towards the absolute calibration of the extragalactic distance scale.
\acknowledgments

WPG and REM gratefully acknowledge support from Fondecyt project 1971076.

\figcaption[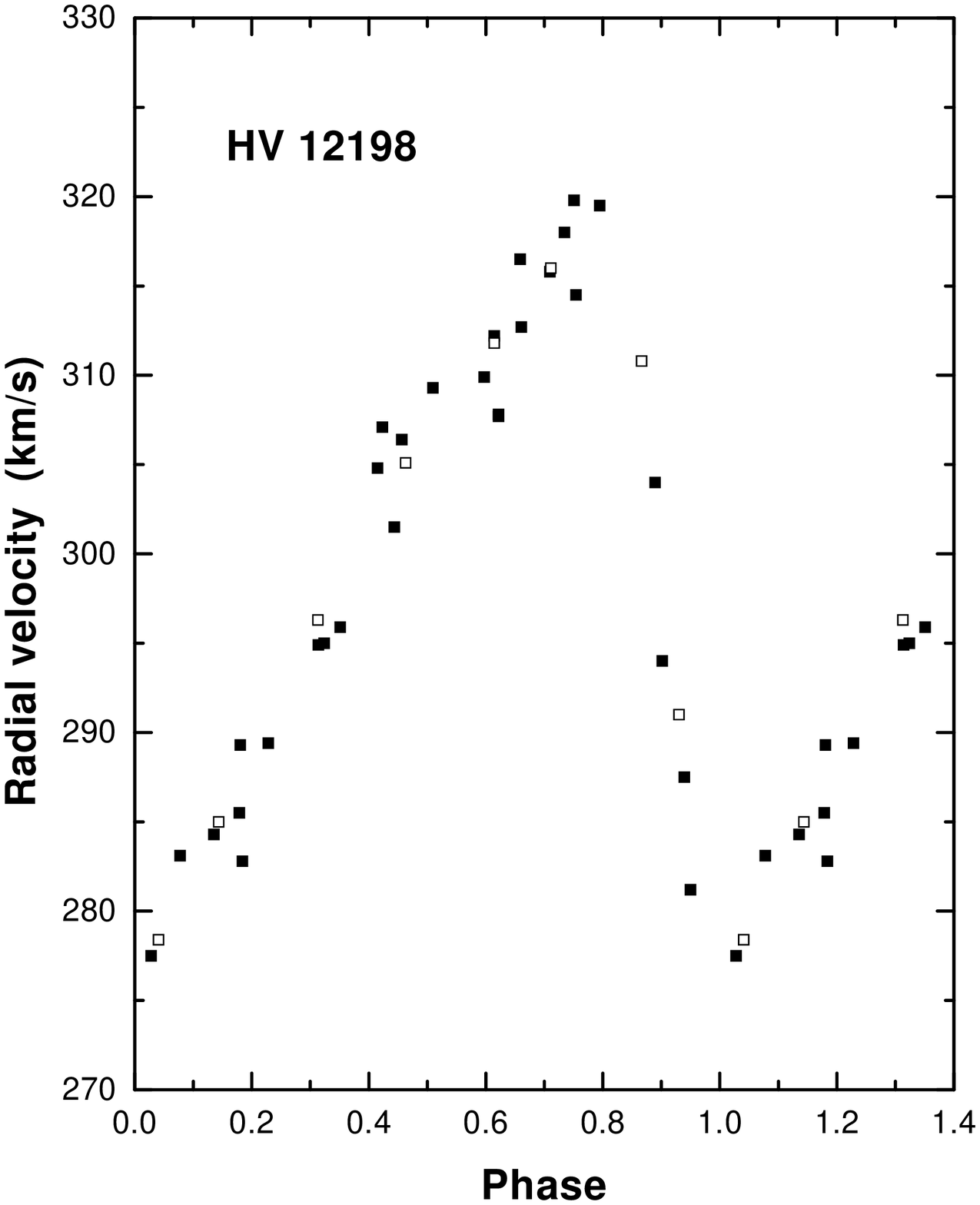]{The radial velocity curve for the Cepheid HV\,12198 in 
NGC\,1866. Filled 
squares, our new observations. Open squares, previous observations of Welch et 
al. 
(1991). The phases have been calculated with a period of 3.52279 days. The mean 
velocity of 
299.8 $\pm$ 1.5 km/s (internal and systematic error) supports cluster membership 
for HV\,12198. There 
is no indication for spectroscopic binarity from the data. \label{fig1}}

\figcaption[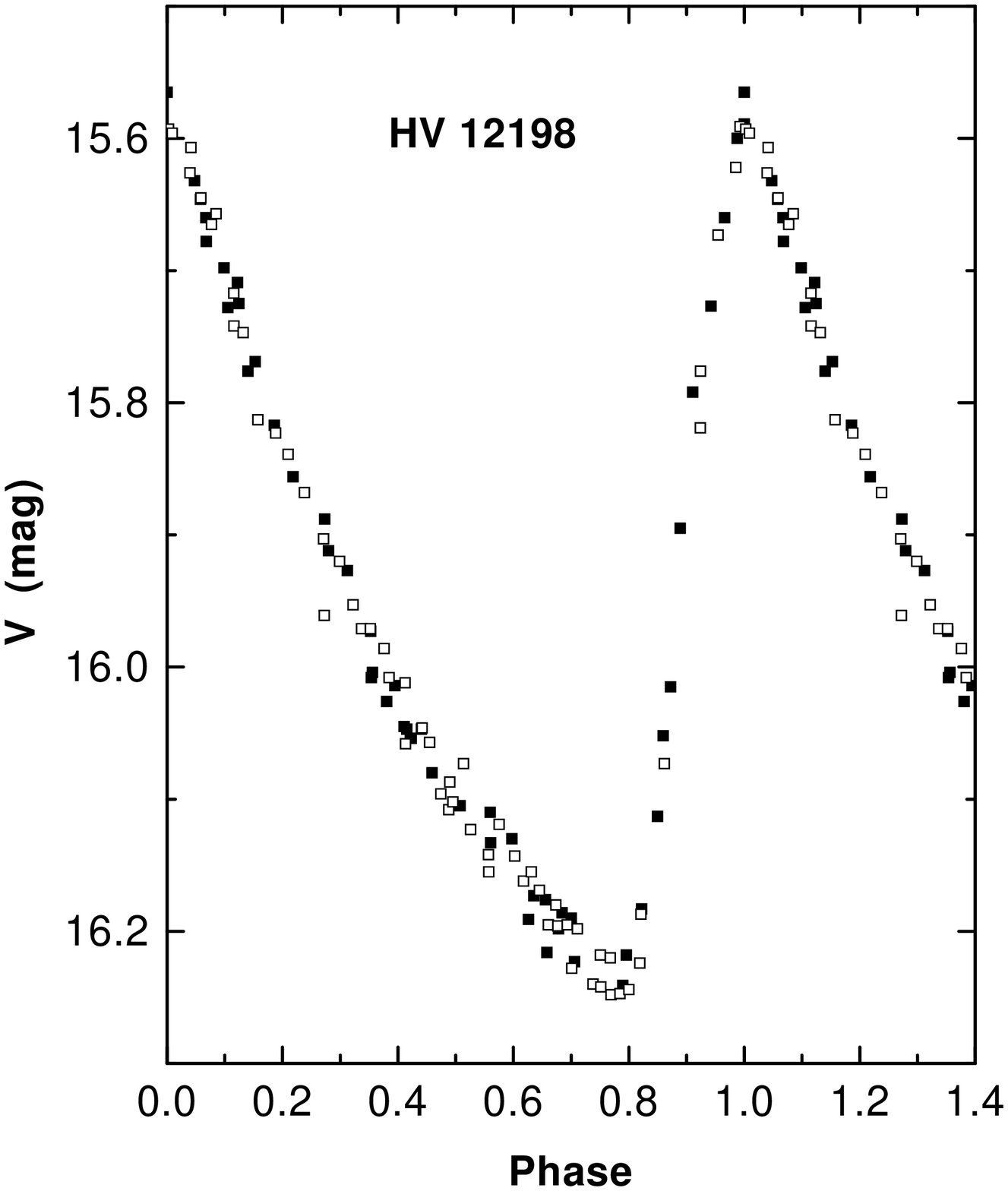]{The V light curve of HV\,12198. Filled squares, our new 
data. Open 
squares, literature data reported by Walker (1987) and Welch et al.\ (1991)  
\label{fig2}}

\figcaption[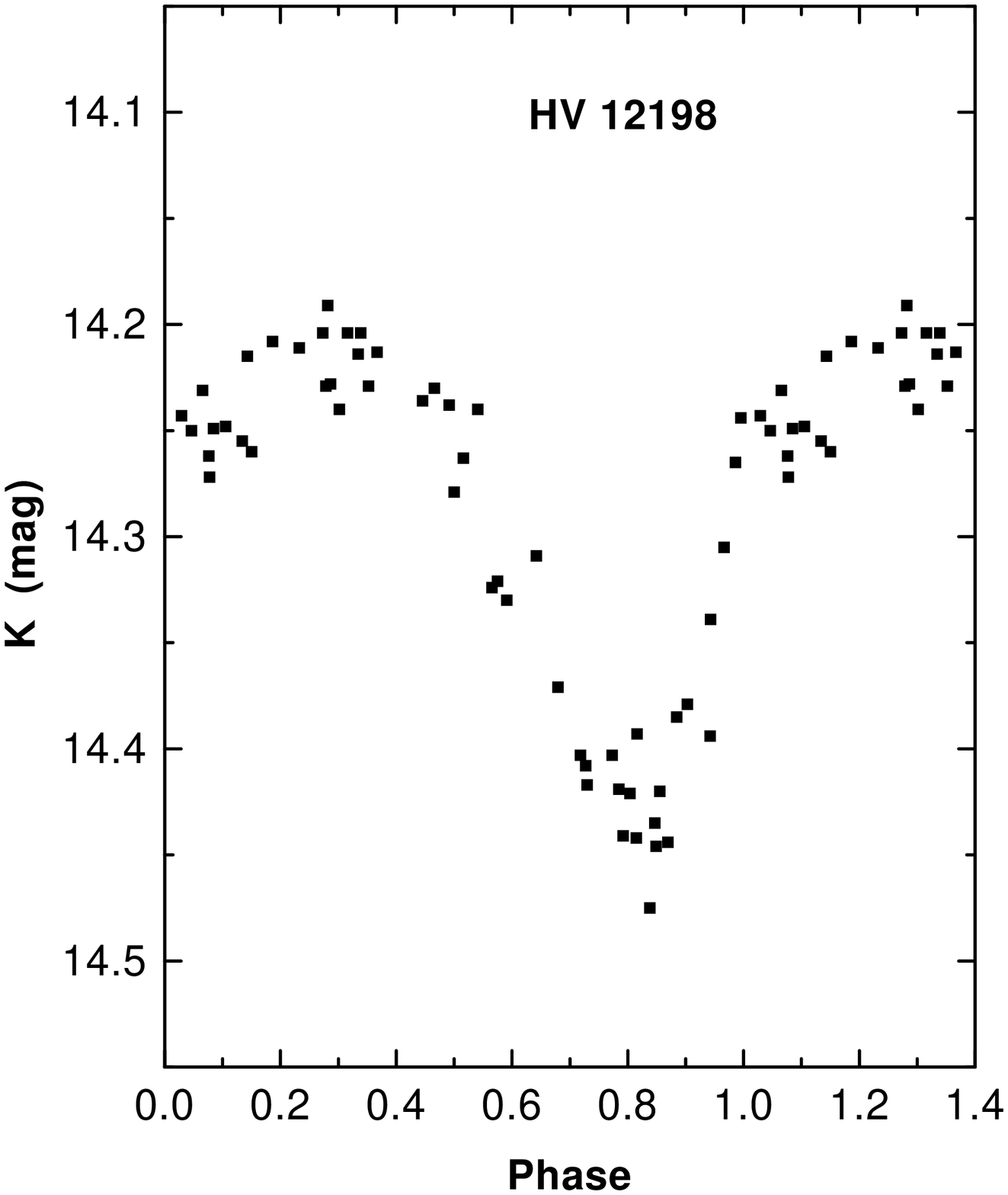]{The K light curve of HV\,12198 from our new data. 
\label{fig3}}

\figcaption[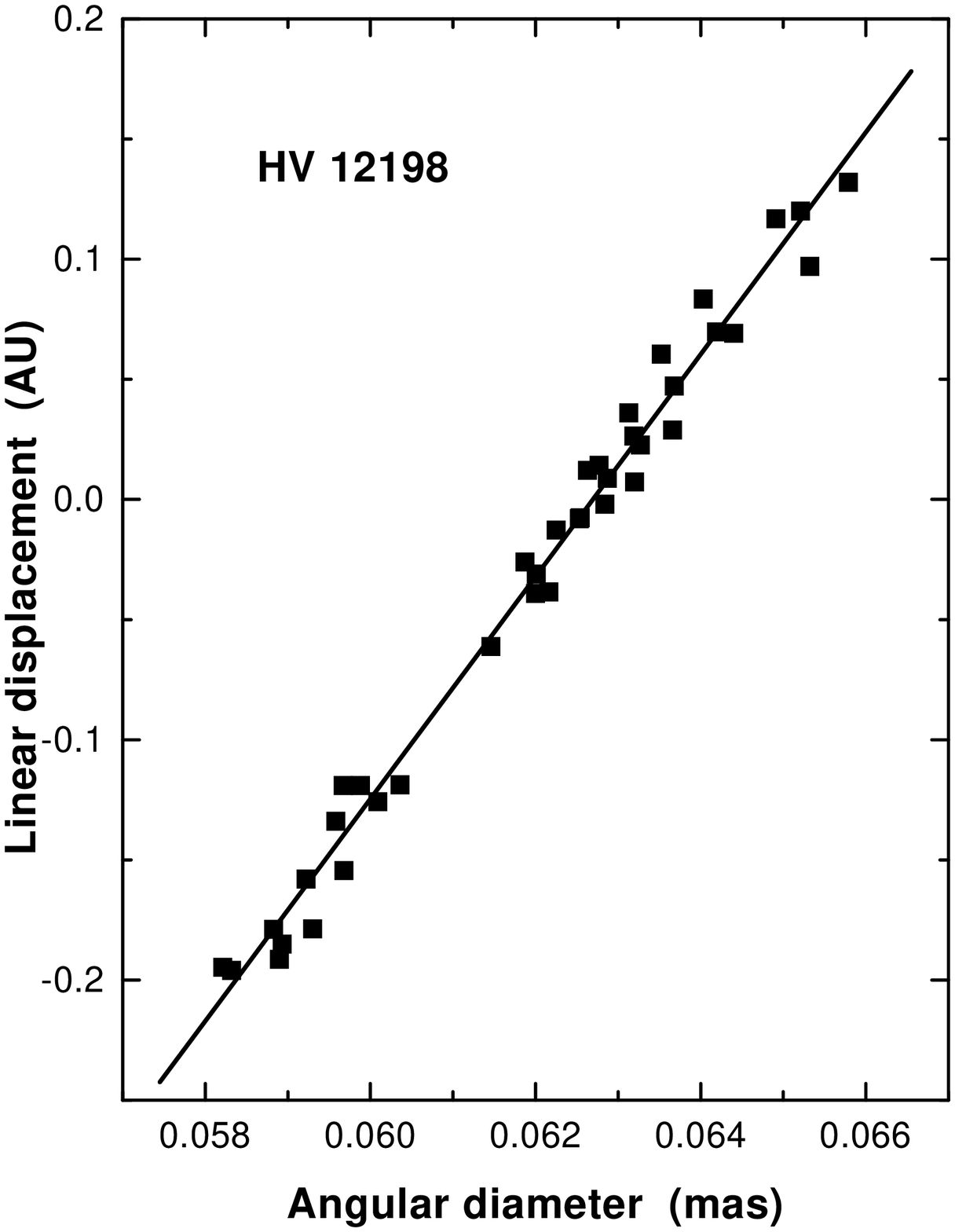]{Linear displacements for HV\,12198, derived from the 
radial velocity 
curve, plotted against the angular diameters at the same phases, derived from 
the VK 
photometry. Solid line is the inverse fit to the data which assumes larger 
relative errors 
on the angular diameters than on the linear displacements. The slope of the line 
yields the 
distance.   \label{fig4}}

\end{document}